\documentstyle[12pt]{article}
\begin{document}
\baselineskip .3in
\begin{center}
{\large{\bf { Nucleon in Nuclear Matter}}} \\
\vskip .2in
 A. Bhattacharya$ ^{\S}$, A. Sagari and B. Chakrabarti  \\
\vskip .1in
Department of Physics, Jadavpur University \\
Calcutta 700032, India.\\
\end{center}
\vskip .3in {\centerline{\bf Abstract}}

The modification of the properties of nucleon in nuclear medium
have been investigated in the context of flux tube model. A
nucleon has been described as diquark-quark system connected by
flux tube and quasi particle diquark model has been used to
describe the diquak constituting the nucleon. The modification of
incompressiblity, the roper resonance etc in the nuclear medium
have been investigated. The results are compared with recent
experimental and theoretical predictions.

 \vskip .3in



\vskip .2in

$\S$e- mail: $pampa@phys.jdvu.ac.in$

\newpage

\vskip 0.4in

EMC effect [1] hints at the modification of the properties of the
nucleon in the nuclear environment particularly swelling of
nucleon nucleus. The experiment shows that the effect is more as
the nuclei become larger and denser. The swelling has largely been
interpreted as the effect of  attraction of the neighbouring
quarks to the quarks constituting the nucleon. This in turn
predicts that there are fewer high momentum quarks in the nucleon
when it is in nuclear matter as swelling means larger wavelength
and small momentum. Recently Seely et al [2] have reported that
the modification of the particle properties do not depend on the
mass or density of the nucleus as a whole but only on the
immediate neighbourhood within nucleus i,e the EMC effect depends
on the local environment. A number of works have been done in the
context of different of models with varied dynamical theory to
explain the EMC effect as the swelling of nucleon in the nuclear
medium[3]. Oka et al [4] have shown from the general ground that a
composite quantum system will swell(shrink) in an external
attractive (repulsive) potential and have concluded that the
nucleon in a nucleus swells due to the attraction of the nuclear
force. Dukelsky et al [5] have investigated the EMC effect
incorporating a mean field acting on the valence quark of nucleon
embedded in nuclear medium. They have obtained a swelling of 2
percent which is much lower than the other estimates [4,6]. Kim et
al [7] have investigated the modification of nucleon form factor
in nucleus. They have estimated anomalous magnetic moment from the
modified form factor and found that it contributes more to the
transverse structure function than the longitudinal one. Wu et al
[8] have studied the modification of the nucleon properties in the
Quark Meson Coupling model(QMC) [9] which includes the effect of
the motion of the quark inside the nucleus. They have studied the
effective mass, radius, correlated charge density disrtibution of
a nucleon in nuclear matter and have observed that the quark
structure plays a crucial role in the description of the nuclear
matter and finite nuclei. Mathieu et al [10] have studied the
modification of hadron properties in the context of the flux tube
model. They have suggested that nuclear binding and EMC effect are
the consequences of the partial deconfienment.

     In the present work we have investigated the modification of
     the properties of a nucleon in nuclear medium in the context
     of the flux tube model. A diquark-quark configuration connected by a flux tube
      has been considered for a nucleon. The quasi-particle diquark
      model has been used to describe the diquark mass.
     The compressibility , Roper resonance, critical density,
     swelling of the nucleon in the nuclear medium have been
     investigated. Some interesting observations are made.
     \vskip 0.4in

In the flux tube model a nucleon is described as a quark and
     a diquark system connected by a flux tube [10]. The Hamiltonian
     for an isolated nucleon can be written as,
\begin{equation}
H = \frac{p^{2}}{2m} + \sigma r -\frac{4}{3} \frac{\alpha_{s}}{r}
\end{equation}
where 'm' is the reduced mass of the system, p is the relative
momentum $\sigma$ and $\alpha_{s}$ are the string tension and the
strong coupling constant respectively. When the nucleon is in
nuclear matter the Hamiltonian changes due to the effect of the
neighbouring nucleons. In the presence of other nucleon the flux
tubes are topologically rearranged such a way that the linear
potential of the system is minimum. Considering the effect of one
perturbing nucleon, the linear potential of the nucleon can be
modified to [10]:
\begin{equation}
V_{Leff}(r) = \sigma r- \frac{\sigma
\rho}{6}\frac{\pi}{24}\int|\psi(r)|^{2}(t^{2}-4u^{2})^{2}\theta(t-2u)d^{3}r'
\end{equation}

where r and r$^{\prime}$ are the variables that represent the
length of the flux tube for the two-nucleon clusterings, t and u
are functions of r and r$^{\prime}$ as in Mathieu et al [10].
$\psi(r)$ is typical nucleon wave function, $\rho$ is the nuclear
matter density, $\sigma$ is string constant.
 The one gluon exchange term also would be modified partially due
 to the presence of the other nucleon and may be represented as in [10]:
 \begin{equation}
 V_{c eff}(r) = -\frac{4}{3}\frac{\alpha_{s}}{r}-\frac{4 \alpha_{s} \rho}{18}\int|\psi(r^{\prime})|^{2}\frac{2\pi}{3}
 (1-Z)(r^{2}-r'^{2}-4rr')d^{3}r'
 \end{equation}
 where z = cos(\textbf{R},\textbf{u}),here R is the distance between two
 nucleons [10].
  The
Hamiltonian can thus be expressed as:
\begin{equation}
 H = \frac{p^{2}}{2m} + V_{L eff}(r) + V_{C eff}
 \end{equation}

 In general the contribution to the potential runs as:
\begin{equation}
 V = \int V_{eff}(r)| \psi(r^{\prime})|^{2}d^{3}r'
 \end{equation}
  In computing the Hamiltonian of the nucleon the
  quark-diquark configuration for a nucleon has been used. For diquark we
  have used the quasi-particle model suggested by us to estimate the
  diquark mass. With diquark within, baryon can
be described as two-body (involving q-qq) system. In the present
study we have considered the proton as a [ud]$_{0}$ (scalar)
diquark and a u quark system and investigated the modification of
the properties of the proton in the nuclear environment. \vskip
0.4in

Recently we have proposed a quasi-particle picture of the diquark
[11] in analogy to the quasi-particle in usual condensed matter
physics. It is well known that the quasi-particles are particle
like entities
     arising in some system of interacting particles. They are low-lying excited states possessing the energy very
     close to the ground state and the properties of the system
     can be obtained by investigating the properties of the quasi-particles to a considerable extent. Such states are often
     observed in the usual condensed matter physics. A crystal electron (with mass m) is subjected to
     two type of forces namely the
effect of crystal field ($\bf{\nabla} V$) and an external force
(\textbf{F}) which accelerates the electron. Under the influence
of the two forces, an electron in a crystal behaves like a
quasi-particle whose effective mass $m^{*}$  reflects the inertia
of electrons which are already in a crystal field [12].
\begin{equation}
\frac{m}{m^{*}}=1 - \frac{1}{\textbf{F}}\frac{\delta V}{\delta r}
\end{equation}

 We assume similar type
of picture for the diquarks. Diquarks are supposed to be
fundamental entities and we have considered that their masses get
modified due to the two types of forces inside the hadron. One is
due to the presence of a background quark field inside the hadron
resembling one gluon exchange type interaction and for external
force harmonic oscillator type of confinement interaction has been
considered. In the context of the potential model potential has
been represented as:
\begin{equation}
V_{ij} = - \frac{\alpha}{r} + (\textbf{F}_{i}.
\textbf{F}_{j})(-\frac{1}{2}Kr^{2})
\end{equation}

where $ \alpha$ is the coupling constant, $ \textbf{F}_{i}.
\textbf{F}_{j}$ = $- \frac{2}{3}$ for qq interaction [13] and K is
the strength parameter, so that,

\begin{equation}
V_{ij} = - \frac{\alpha}{r} + ar^{2}
\end{equation}

where a = K/3. The mass of the diquark in the background of the
quark field gets modified in the quasi particle approximation as
it happens in the crystal lattice [12].
   While retrieving the  $\frac{1}{\textbf{F}}\frac{\delta V}{
\delta r}$ part we have considered the one-gluon exchange
potential for background field and for external force F we have
used the confinement force [11]. Thus we come across:

\begin{equation}
\frac{m_{q_{i}}+m_{q_{j}}}{m_{D}}=1 + \frac{\alpha}{2ar_{D}^{3}}
\end{equation}

With $m_{u}$= $m_{d}$ = 360 MeV, $\alpha$ =
$\frac{2}{3}\alpha_{s}$ =0.393 as $\alpha_{s}$ = 0.58, K = 241.5
MeV$fm^{-2}$ [13] the diquarks masses have been estimated as
$[m_{ud}]$= 506 MeV. We have used the wave function for the baryon
as suggested in the context of the Statistical Model [14]. In this
model, a hadron is assumed to be
 consists of a virtual $q\overline{q}$ in addition to the valence
 partners which determines the quantum number of the colourless
 hadron. The quarks, real and virtual, are assumed to be of same colour and flavour
 so that they may be regarded as identical and indistinguishable and are treated on the same
 footing. The indistinguishability of the valence quark with the
 virtual partner calls for the existence of quantum mechanical
 uncertainty in its available phase space. The valence quarks are
 assumed to be non-interacting with each other and
 considered to be moving almost independently in conformity with the
 experimental finding of asymptotic freedom. However it is considered that the valence
 quarks move in an average smooth background
 potential due to their interaction with the virtual sea.
 With the above consideration we have arrived at an expression
 for the probability density of the baryon in the ground state as [14],
 \begin{equation}
 |\phi(r)|^{2} = \frac{315}{64\pi
 r_{0}^{9/2}}(r_{0}-r)^{3/2}\theta(r_{0}-r)
 \end{equation}

 where r$_{0}$ is the radius parameter of the corresponding hadron and
 $\theta(r_{0}-r)$ represents the usual step function. In the modified version of the model, the
 exchange effect which arise from the tendency of quarks of like spin to keep them apart
 as suggested by Dirac [15] has been incorporated. The wave function is thus obtained as [16],
\begin{equation}
 |\phi(r)|^{2} = A(r_{0}-r)^ \frac{3\mu_{L}}{2}\theta(r_{0}-r)
 \end{equation}
 The exchange effect has been incorporated through a term $\mu_{L}$ =
$\frac{\alpha_{ex}}{a}(\frac{dp}{dr})$.
  The normalization constant is obtained as A = [$4 \pi
 r_{0}^{\frac{3\mu_{L}}{2}+3}B(3,\frac{3\mu_{L}}{2}+1)]^{-1}$
 where B is the beta function.
  Using these wave functions as input in the expression (2) and (3), and taking $\mu_{L}$ = 3.66 [17] the Hamiltonian are obtained as;
  \begin{equation}
  <H> = \frac{28.079}{r_{0}^{2}}+ 0.545\sigma r_{0} -0.055\sigma
  \rho r_{0}^{4}-\frac{3\alpha_{s}}{r_{0}}+0.312\alpha_{s}\rho r_{0}^{2}
  \end{equation}
  and
\begin{equation}
  <H> = \frac{46.205}{r_{0}^{2}}+ 0.316\sigma r_{0} -0.0088\sigma
  \rho r_{0}^{4}-\frac{5.66\alpha_{s}}{r_{0}}+ 0.112\alpha_{s}\rho r_{0}^{2}
  \end{equation}
  Neglecting the Coulomb term and minimizing the Hamiltonian with
  respect to radius parameter we come across the following expressions corresponding to equations (12)
  and (13) respectively as:
  \begin{equation}
  r_{0}=(\sigma/206.048]^{-\frac{1}{3}}[1\pm(1-106.516\rho/\sigma)^{\frac{1}{2}}]^{-\frac{1}{3}}
  \end{equation}
  and
 \begin{equation}
  r_{0}=(\sigma/584.88]^{-\frac{1}{3}}[1\pm(1-130.19\rho/\sigma)^{\frac{1}{2}}]^{-\frac{1}{3}}
  \end{equation}
  The compressibility of a nucleon can be expressed as:
 \begin{equation}
 K = \frac{1}{3}r_{0}^{2}(\frac{d^{2}E}{dr^{2}})
 \end{equation}
 The expression for Roper excitation energy runs as;
\begin{equation}
 \Delta E = (\frac{K}{m_{q}r_{0}^{2}})^{\frac{1}{2}}
 \end{equation}

   From the expressions (14) and (15) it has been found that $\rho$ = $\sigma$/106.516 and $\rho$ = $\sigma$/130.19
   respectively which corresponds to the real root of the radius
   parameters and can be defined as the de-confining density or
    the critical density $\rho_{c}$ of the medium at which the
    nucleon ceases to exist. We have estimated the compressibility, Roper
    Resonance with the input of the radii obtained with string constant $\sigma$ = 0.219
  GeV$^{2}$ [18] from equations (14) and (15) corresponding to $\rho$ = 0. The results are displayed in Table-I and
  Table-II corresponding to the energy expressions from equations
  (12) and (13).

\vskip .4in

Table I: Compressibility, Roper Resonance Binding Energy, critical
density with $|\phi(r)|^{2}$ from [10]
 \vskip 0.4in
\begin{tabular}{|r|r r|r|r|}
\hline

$  $&$Compressibility$&$ $&$  $&$  $\\

 $\sigma$&$K_{f}$&$K_{b}$& $Roper Resonance$ &$\rho_{c}$\\
 $ (GeV^{2}) $&$(GeV)$&$(GeV)$&$B.E.(GeV)$&$  $\\

\hline

 0.219&0.787&0.562&0.029&1.51 $\rho_{0}$\\

\hline
\end{tabular}

\vskip 0.4in

 \vskip .4in

Table II: Compressibility, Roper Resonance Binding energy,
critical density with $|\phi(r)|^{2}$ from [10]
 \vskip 0.4in
\begin{tabular}{|r|r|r r|r|r|}
\hline

$  $&$Compressibility$&$ $&$  $&$  $\\

 $\sigma$&$K_{f}$&$K_{b}$& $Roper Resonance$ &$\rho_{c}$\\
 $ (GeV^{2}) $&$(GeV)$&$(GeV)$&$B.E.(GeV)$&$  $\\

\hline

 0.219&0.635&0.517&0.012&1.24 $\rho_{0}$\\

\hline
\end{tabular}

\newpage
\textbf{Conclusions:} \vskip.2in
  In the present work we have investigated the properties of the
  nucleon in the nuclear matter in the context of the flux tube
  model considering a nucleon consist of a quark and a diquark
  connected by a flux tube. Quasi-particle diquark model [14] is used for describing the diquark. The compressibility, Roper excitation energy
  have been estimated. The results obtained are found to be in the reasonable agreement with the existing predictions.
   The nucleon shows a
  swelling in the nuclear medium
  and the
  percentage increase in radius at the critical density have been
  estimated to be of 26 percent. In the context of the non-topological soliton
  model Wen et al [19] have estimated the increase in nucleon
  radius by 10 to 16 percent at normal nuclear matter density
  whereas Noble [20] has observed the increase in charge radius of nucleon to be 30
  percent.
  Mathieu et al [10] have observed 25 percent increase in
  radius at deconfinement density in the context of the flux tube
  model. Dukelsky et al [5] have obtained
  the swelling to be of 2 percent much lower than the existing
  predictions mentioning the limitation of the model they have used.
  Chang et al [21] have investigated the
  modification of nucleon properties in the context of the Global
  Colour Symmetry Model and have observed that the nuclear
  density is 8 times the normal nuclear density ($\rho_{0}$= 0.17 fm$^{-3}$) at which the
  radius becomes infinite representing a phase transition.
  Achtzehnter et al [22] obtained the critical density to be of the
  order of normal nuclear density in the context of soliton bag model
   and argued that the model cannot adequately incorporate the
   repulsive part of the nucleon-nucleon interaction at small
   relative distance. The critical density obtained in the present work
  seems to be lower than the other existing estimates.
   The difference in  compressibility ($\Delta$K) between the free nucleon and
   nucleon in the medium at normal nuclear density ($\rho_{0}$= 0.17 fm$^{-3}$) has been found to be 0.225 GeV and
   0.118 GeV for two version of the Statistical Model used. The reduction of compressibility in medium is found to be
     substantial
     in the present investigation.
     The decrease in compressibility of the nucleon
   the medium shows that the nucleons in nuclear medium are more diffused than free
   nucleons. The Roper resonance excitation energy is
    more in free nucleon than when it is in nuclear
   medium. The difference have been obtained as 0.029 GeV and 0.012
   GeV in the present work. Mathieu et al [10] obtained the value
   as 0.021 GeV whereas the most accepted value is $\sim$0.040 GeV.

   In the current investigation we have neglected the contribution from Coulomb
   term. It may be mentioned that the most uncertainty comes from the radius parameter of the nucleon
    which has a wide range of values in the literature.
   The string constant $\sigma$ also have important role
    in determining the in medium properties and
    precise knowledge of its value is yet to be determined. It has a range of values [18]. It is
    interesting to observe that in the current work we have come
    across an expression where the radius parameter is related to
    the string constant. If we put radius of the proton to be
    0.865 fm which is the charge radius of the proton and suggested
    to be the equal to the radius parameter defining it as a meso
    object [23], the string constant would have the value
    1.26 GeV$^{2}$ which is quite high compared to the common accepted
    values. Compressibility also has wide range of values and is
   under
   discussion in the current literature.
   It has been suggested that the information on the compressibility of a system can be obtained
   from the dynamical properties of the size degree of
   freedom [24]. In the current
   work
   the quark-diquark model (quasi-particle diquark model) of the nucleon has been
   used to study the in medium modification of the nucleon
   properties in the context of the Statistical Model both in
   original and modified version. The exchange effect of the quark
   spin [15] seems to reduce values of all the properties yielding
    comparatively
    larger
    value of the radius parameter.
\vskip.1in In the present investigation
   the modification of the nucleon properties have been done
    taking into the consideration of medium effect through the
    density of the nuclear matter. However it may be pointed out
    that recently it has been suggested by Seely et al [2] that nucleon dependence of quark distribution of nucleon depends on the local environment not on A or
   the density. More experimental and theoretical efforts are needed to reveal
    the in medium effect on the nucleon which in turn would help us to know exact nature of the underlying
    interactions.

\vskip 0.4in

\newpage
{\bf References}

\vskip .4in

\noindent [1]. J.J.Aubert et al, Phys.Lett.  {\bf B 123} (1983)
275.

\noindent [2]. J.Seely, Phys. Rev. Letts {\bf 103} (2009) 202301.

\noindent [3]. S.Sarangi et al, arXiv.0805.2449v1[nucl-th](2008);
M.K. Banerjee, Phys.Rev.C{\bf 45} (1992) 1359; A.T.D'yachenko et
al.Phy. Scr.{\bf T 104} (2003) 91 ;B. Desplanques, Euro.Phys.J{\bf
A 330} (1988) 331.

\noindent [4]. M.Oka and R.D.Amado, Phys. Rev. {\bf C 35} (1987)
1586.

\noindent [5]. D.Dukelsky, J. Phys. {\bf G 35}  (1995) 317.

\noindent [4]. M.Oka and R.D.Amado, Phys. Rev. {\bf C 35} (1987)
1586.

\noindent [5]. D.Dukelsky, J. Phys. {\bf G 35}  (1995) 317.

\noindent [6]. K.Braucr et al, Nucl. Phys.{\bf A 347}(1985) 717.

\noindent [7]. K.S. Kim et al, Euro. Phys. J. {\bf A 8} (2000) 131

\noindent [8]. W. Wu and S. Hong, Chinese. J. Phys Phys. {\bf C
32} SupplII (2008) 81.

\noindent [9]. H. Muller and B.K. Jennings, Nucl. Phys.{\bf A 626}
(1997) 966

\noindent [10]. P. Mathieu and P.J.S. Watson, Can. J. Phys {\bf
64} (1986) 1389 .

\noindent [11]. A. Bhattacharya at al, Phys. Rev{\bf C 81}(2010)
015202; B. Chakrabarti et al, Phys. Scrp.{\bf 79}(2009) 025103, B
Chakrabarti et al; Nucl. Phys.{\bf A 782}(Proc. Suppl.), 1-4,
392c, (2007).

\noindent [12].  A. Haug;Theoretical Solid State Physics {\bf Vol
I }( Pergamon Press, Oxford,1975),Pg 100.

\noindent [13]. R K Bhaduri et al, Phys. Rev. Lett. {\bf 44}
(1980)1369.

\noindent [14]. A. Bhattacharya et al, Nucl. Phys.{\bf B 142}
(2005), 13 (Proceedings Suppl);A. Bhattacharya et al, Mod. Phys.
Lett{\bf A 19}(2004) 921; S. N. Banerjee et al,;Ann. Phys
(N.Y.){\bf 150}(1983) 150.

\noindent [15]. P.A.M. Dirac, Proc.Camb.Phil.Soc {\bf 2b}(1930)
376.

\noindent [16]. S. N. Banerjee et al, Int. J. Mod. Phys {\bf
A16}(2001)201, Had.J {\bf 11}(1988) 243.

\noindent [17]. S. N. Banerjee et al, Had. J.{\bf 11} (1988)243;
Had. J.{\bf 12}(1989)178; Had. J.{\bf 13} (1992)75.

\noindent [18]. G.L. Strobel; Int. J. Theor.Phys.{\bf 39} (2000)
2853.

\noindent [19]. W. Wen and H.Shen, Phys. Rev.{\bf 77} (2008)
065204.

\noindent [20]. J.V.Noble, Phys.Rev. Lett. {\bf B46}(1981) 412.

\noindent [21]. L. Chang et al, Nucl. Phys. {\bf A 750}(2005) 324.

\noindent [22]. J. Achtzehnter and W.Scheid, Phys. Rev.{\bf D 32}
(1985) 2414.

\noindent [23]. E.F. Hefter, Phys. Rev. {\bf A 32}(1985) 1201.

\

\noindent [24]. T. Meissner, Phys. Rev. {\bf D 39}(1989) 39.

\end{document}